\input{aipcheck}

\documentclass[
    ,final            % use final for the camera ready runs
	, numbers
  ]
  {aipproc}

\layoutstyle{8x11single}

\usepackage{subfig}
\usepackage{multirow}

\newcommand{\ca}{capacitance }

\def\tect{$^{130}$Te}

\def\udto{$^{238}$U}
\def\tdt{$^{230}$Th}
\def\rdvs{$^{226}$Ra}
\def\rdvd{$^{222}$Rn}
\def\pddo{$^{218}$Po}

\def\pdd{$^{210}$Po}
\def\tdtd{$^{232}$Th}
\def\tdvo{$^{228}$Th}
\def\advo{$^{228}$Ac}
\def\rdvq{$^{224}$Ra}
\def\pds{$^{216}$Po}
\def\tldo{$^{208}$Tl}
\def\kq{$^{40}$K}
\def\cs{$^{60}$Co}

\def\bdq{$^{214}$Bi}
\def\pcn{$^{190}$Pt}
\def\sod{$^{82}$Se}
\def\ccs{$^{116}$Cd}
\def\mc{$^{100}$Mo}

\def\alfa{$\alpha$}

\def\mbb{$\langle m_{\beta\beta} \rangle$}
\def\BBz{$\beta\beta(0\nu)$~}
\def\BBzn{$\beta\beta(0\nu)$}

\def\BBdn{$\beta\beta(2\nu)$}

\def\ca{$\sim$}

\def\dot{$\cdot$}
\def\pom{$\pm$ }
\def\gm{$\gamma$}

\def\teod{TeO$_2$~}

\def\be{\begin{equation}}
\def\ee{\end{equation}}

\def\per{$\times$}

\def\zs{$ZnSe$}
\def\ct{$CdWO_4$}
\def\zm{$ZnMoO_4$}
\def\ckky{counts/(keV\dot kg\dot y)}

\makeatletter
\let\MakeTitle\maketitle % storing LaTeX's or so
    \renewcommand*{\MakeTitle}{\let\@fnsymbol=\@alph \maketitle}
\makeatother

\begin{document}

\title{CUORE-0 results and prospects for the CUORE experiment}
%\date{2005/12/01}


\author{D.~R.~Artusa}{address={Department of Physics and Astronomy, University of South Carolina, Columbia, SC 29208 - USA},
altaddress={INFN - Laboratori Nazionali del Gran Sasso, Assergi (L'Aquila) I-67010 - Italy}}
\author{F.~T.~Avignone~III}{address={Department of Physics and Astronomy, University of South Carolina, Columbia, SC 29208 - USA}}
\author{O.~Azzolini}{address={INFN - Laboratori Nazionali di Legnaro, Legnaro (Padova) I-35020 - Italy}}
\author{M.~Balata}{address={INFN - Laboratori Nazionali del Gran Sasso, Assergi (L'Aquila) I-67010 - Italy}}
\author{T.~I.~Banks}{address={Department of Physics, University of California, Berkeley, CA 94720 - USA},
altaddress={Nuclear Science Division, Lawrence Berkeley National Laboratory, Berkeley, CA 94720 - USA}}
\author{G.~Bari}{address={INFN - Sezione di Bologna, Bologna I-40127 - Italy}}
\author{J.~Beeman}{address={Materials Science Division, Lawrence Berkeley National Laboratory, Berkeley, CA 94720 - USA}}
\author{F.~Bellini}{address={Dipartimento di Fisica, Sapienza Universit\`a di Roma, Roma I-00185 - Italy},
altaddress={INFN - Sezione di Roma, Roma I-00185 - Italy}}
\author{A.~Bersani}{address={INFN - Sezione di Genova, Genova I-16146 - Italy}}
\author{M.~Biassoni}{address={Dipartimento di Fisica, Universit\`a di Milano-Bicocca, Milano I-20126 - Italy},
altaddress={INFN - Sezione di Milano Bicocca, Milano I-20126 - Italy}}
% ...

\begin{abstract}

With 741 kg of \teod crystals and an excellent energy resolution of 5 keV (0.2\%) at the region of interest, 
the CUORE (Cryogenic Underground Observatory for Rare Events) experiment aims at searching for neutrinoless double beta decay of \tect~ with unprecedented sensitivity.
Expected to start data taking in 2015, CUORE is currently in an advanced construction phase at LNGS. 
CUORE projected neutrinoless double beta decay half-life sensitivity is $1.6\times 10^{26}$~y at $1\sigma$ ($9.5\times10^{25}$~y at the 90~\% confidence level), in five years of live time, corresponding to an upper limit on the effective Majorana mass in the range 40--100~meV (50--130~meV). 
Further background rejection with auxiliary bolometric detectors could improve CUORE sensitivity and competitiveness of bolometric detectors towards a full analysis of the inverted neutrino mass hierarchy. 
CUORE-0 was built to test and demonstrate the performance of the upcoming CUORE experiment. 
It consists of a single CUORE tower (52 \teod bolometers of 750 g each, arranged in a 13 floor structure) constructed strictly following CUORE recipes both for materials and assembly procedures.
An experiment its own, CUORE-0 is expected to reach a sensitivity to the \BBz half-life of \tect~ around 3\per 10$^{24}$ y in one year of live time. 
We present an update of the data, corresponding to an exposure of 18.1 kg y. 
An analysis of the background indicates that the CUORE performance goal is satisfied while the sensitivity goal is within reach.

\end{abstract}

\MakeTitle
%\maketitle

%\tableofcontents

\bigskip

\section{1. Introduction}
The question wether neutrinos are their own anti-particles (i.e. wether they are Majorana fermions) is among the most fundamental open questions of particle physics \cite{bil12}.
Proposed more than 70 years ago to establish the nature of neutrinos \cite{fur39},  neutrinoless double beta decay is still the most sensitive probe into the non-conservation of lepton number. 
Its observation would be a breakthrough in our description of elementary particles and would provide fundamental information on the neutrino masses and their origin.
In fact, double beta decays are a family of very rare nuclear processes in which a nucleus transforms into one of its isobars with a change of the nuclear charge by two units while emitting two electrons.
The two neutrino mode (\BBdn) in which two daughter neutrinos are also emitted $(A,Z) \to (A,Z+2) + 2e^- +2\overline{\nu}$, is a 2$^{nd}$-order weak process allowed by the Standard Model and observed in a number isotopes. 
On the contrary, the neutrinoless mode (\BBzn) $(A,Z) \to (A,Z+2) + 2e^-$, can occur only if neutrinos are Majorana particles and is by far the most realistic lepton number violating process. Neutrino masses can account for the chirality adjustment needed to take into account for parity violation in weak interactions, thus \BBz becomes also a powerful tool to provide informations about the absolute scale of neutrino masses and possible CP violations on the lepton sector.
For \BBz with the exchange of a light Majorana neutrino, the decay rate $\Gamma$ can be expressed as
\begin{equation}
	\Gamma = G^{0\nu} \vert M^{0\nu}\vert ^2 {{\vert\langle m_{\beta\beta}\rangle\vert^2}\over{m_e^2}}
\end{equation}

where \mbb = $\sum_{k}U_{ek}^2 m_j$ is known as ``effective Majorana mass'', and is a combination of the neutrino mass eigenvalues weighted on the elements of the PMNS (Pontecorvo-Maki-Nakagawa-Sakata) mixing matrix, $G^{0\nu}$ is the phase space factor and  $M^{0\nu}$ is the nuclear matrix element (NME) of the transition.

$G^{0\nu}$  is proportional to the 5$^{th}$ order of decay Q-value and can be accurately calculated \cite{kot13}. On the contrary, NME's evaluation is difficult and different model calculations may disagree by a factor of 2 to 3, thus introducing a large spread of \mbb \cite{nmes}. 

All \BBz experiments measure the energy of the two emitting electrons. The experimental signature is thus a sharp peak at the decay Q-value in the summed electron energy spectrum. 
Currently, more than 10 experiments world-wide aim at searching for \BBz utilizing different techniques such as ionization, scintillation, thermal excitation, or some combination of them. 

For all the next generation \BBzn, very stringent requirements on the parent isotope mass, detector energy resolution, and background rate must be satisfied. 
For the background-limited case, the commonly accepted figure of merit for the half-life limit is

\begin{equation}
	T^{0\nu}_{1/2} \propto \eta \cdot \epsilon \sqrt{{M \cdot t}\over{b\cdot\Delta E}}
\end{equation}

where $\epsilon$ is the detector efficiency, $\eta$ the isotopic abundance of the \BBz candidate, M the total detector mass, t the detector live time, b the background rate per unit detector mass per energy interval, and $\Delta$E is the energy resolution of the detector. When the background index is so low that no background counts ($N_b = b\cdot\Delta E\cdot t\cdot M$) are expected along the measurement time (a condition usually known as ``zero background''), a very favorable condition occurs where the half-life limit scales linearly with the exposure M\dot~t and is independent of the background rate and the energy resolution:

\begin{equation}
	T^{0\nu-ZB}_{1/2} \propto \eta \cdot \epsilon \cdot M \cdot t
\end{equation}

\BBz process has never been observed, and current experimental sensitivities are in the range 10$^{24-25}$ years \cite{cre13}.

The CUORE experiment \cite{arn04,art14} will search for \BBz decay of \tect~ using a large array of $^{nat}$\teod cryogenic bolometers acting at the same time as source and detector of the decay. 
%Typically we categorize searches for \BBz into two groups by the relation of \BBz source and detector. One common approach is to implement a detector with candidate isotope inside.  For example, GERDA uses semiconductor 76Ge-enriched germanium diode to search for \BBz of \ges \cite{ago13}. The homogeneous (or ``source = detector'') approach has the advantage of a high detecting efficiency.  The other approach emphasizes on the flexibility of \BBz source choices by separating the source from detection, the typical example of which is the NEMO3/SuperNEMO project \cite{bar11}. The bolometric technique combines the benefits of high efficiency and flexibility of candidate isotope choice. For large-mass bolometers such as CUORE modules, the probability of both \BBz electrons are confined in the crystal is (87.4 \pom 1.1)~\%.  The bolometric technique puts little constraints on the absorber materials, other than preferably dielectric for the benefit of small heat capacity. Crystals such as \teod or ZnSe are easily interchangeable for search of \BBz in different candidate isotopes. The flexibility is especially useful for confirming \BBz discoveries or cross-comparing m limits in different isotopes.
Indeed, \tect~ is an attractive choice as \BBz emitter: its natural isotopic abundance (34.2\%) is the highest among all the \BBz candidate isotopes \cite{feh04}, and therefore no isotopic enrichment is necessary, and the decay Q-value is measured to be around 2528 keV \cite{qval}, which is higher than most naturally occurring \gm-ray background and also means a favorable phase space factor.
The CUORE bolometers will operate at a temperature of about 10 mK which allows them to achieve an average energy resolution of about 5 keV FWHM at the Q-value of
the decay. 
CUORE is currently in an advanced phase of construction at Laboratory Nazionali del Gran Sasso (LNGS) in Italy, and is anticipated to start data taking in 2015.
The technology of CUORE was proven by its predecessor, Cuoricino, a \ca 40 kg tower of 62 bolometers that, with 19.75 kg y of \tect~ exposure, set a lower limit to the decay half-life of  2.8 \per 10$^{24}$ y (90\% C.L.) \cite{and11}.

Cuoricino background rate in the region of interest (ROI) around the \BBz Q-value was dominated by the contribution of surface radioactive contaminations of the detector structure. 
Therefore, to improve the background for CUORE, we implemented a set of strict protocols to limit the radioactive contamination of the detector materials during production and assembly.
The first tower produced on the CUORE detector assembly line and operated separately in the same cryostat which  hosted Cuoricino, has been called CUORE-0 \cite{agu14}.
It has been taking data since March 2013. 
CUORE-0 is a crucial test of the CUORE concept in all stages as well as an independent experiment in its own, expected to surpass the Cuoricino sensitivity in one year of live time. 

In this article, we give an update on latest CUORE progress and CUORE-0 results corresponding to a 18.1 kg\dot~y \teod exposure.
The Cuoricino background model, confirmed by CUORE-0 data, anticipates that CUORE sensitivity will be limited by background, most of which originates from the surface radioactive contaminations of detector materials and nearby structures. 
Future generations of bolometer arrays focus therefore on background suppression by the simultaneous detection of photons and phonons. 
The photon signal comes either from Cherenkov radiation or scintillation light and can be applied to the most interesting \BBz candidates \cite{ihe} as discussed in section 4. %\ref{sec:ihe}.

\section{2. CUORE}
The CUORE concept is based on the low temperature calorimeter technology, in which the energy release in an absorber is measured via its temperature rise, picked up by a sensitive thermal sensor attached to the absorber.
This requires to operate at cryogenic temperatures (\ca 10~mK) in order to minimize the detector heat capacity. 
The intrinsic sensitivity and resolution of bolometers are then excellent and, when coupled with an unparalleled free choice of the  materials (the only request for the absorber is usually just to be dielectric and diamagnetic), make this technique ideal for applications in the Physics of rare events \cite{fio84}.

%\begin{figure}
%    \includegraphics[height=0.26\textheight]{pics/CUORE_detector}
%\hspace{20pt}
%    \includegraphics[height=0.26\textheight]{pics/CUORE0_detector}
%\hspace{40pt}
%    \includegraphics[height=0.26\textheight]{pics/CUORE0_picture}
%    \caption{Schemes of the CUORE and CUORE-0 detectors and a picture of the CUORE-0 tower.}
%    \label{fig:cuoredet}
%\end{figure}

The CUORE detector is a large bolometer array consisting of 988 independent \teod crystal modules arranged in 19 towers.
% (Figure \ref{fig:cuoredet} left). 
Each tower has 13 floors with 4 bolometers on each floor in a two by two configuration. 
The main component of each module is a 750 g, 5-cm cubic \teod crystal grown from natural tellurium. The crystals were manufactured by the Shanghai Institute of Ceramics, Chinese Academy of Sciences with major input from the CUORE collaboration on quality and radio purity control \cite{cuorecrystals}. 
The total detector mass is 741 kg, corresponding to 206 kg of the \tect~ isotope. 
3\per 3\per 1 mm$^3$ NTD Ge thermistors \cite{ito94} have been glued  on each module as temperature sensors.
Each bolometer is also instrumented with a Joule heater, which is used to inject in each bolometer a known amount of energy at regular time intervals for offline thermal gain instability correction \cite{fan11}.
Test modules operated underground in a dedicated cryogenic test facility have been featuring good reproducibility, excellent stability and low noise. 
The design full width at half maximum (FWHM) resolution at ROI is 5 keV. Energy resolutions of large-mass bolometers are dominated by thermal fluctuations from vibration and readout electronics. 
After production, the crystals were transported to LNGS at sea level to minimize the cosmogenic activation. We performed dedicated cryogenic tests to measure the bulk and surface contamination rates and determined them to be less than 6.7\per 10$^{-7}$ Bq/kg and 8.9 \per 10$^{-9}$ Bq/cm$^2$ at 90~\% C.L. in \udto, respectively, and less than 8.4 \per 10$^{-7}$ Bq/kg and 2.0 \per 10$^{-9}$ Bq/cm$^2$ in \tdtd, respectively \cite{validationruns}. 
To mitigate the surface contamination of the copper structure, we tested three surface treatment techniques, and chose a series of tumbling, electropolishing, chemical etching, and magnetron plasma etching for the surface treatment. The upper limit of the surface contamination of the cleaned copper was measured in R\&D bolometers to be 1.3 \per 10$^{-7}$ Bq/cm$^2$ at 90~\% C.L. in both \udto~ and \tdtd~  \cite{ale12}.
CUORE-0 results (sec.~3) show that this goal has been achieved.

The projected background rate at ROI is 0.01 \ckky, mainly from surface and bulk contaminations of the bolometer components and the innermost copper thermal shields. 
The best estimate for surface-related background, mainly from \alfa~particles originated from the copper surface, is 1-2 \per 10$^{-2}$ \ckky, based on Monte Carlo simulations with the latest results of CUORE-0 \cite{art14b}. 
Background contribution from bulk contamination, predominantly events from \tldo, is expected to be less than 6 \per 10$^{-3}$ \ckky (90~\% C.L.) \cite{art14b}.

The CUORE building is situated at the LNGS underground facility at an average depth of 3650 m water equivalent, where the muon flux is (2.58 \pom 0.3) \per 10$^{-8}$ $\mu /(s \cdot cm^2)$, about six orders of magnitude smaller with respect to sea level \cite{mei05}. 

A heavy shield consisting of layers of borated polyethylene, boric-acid powder, and lead bricks surrounds the cryostat to attenuate neutron and \gm-ray backgrounds. 
More lead shielding is added inside the cryostat, including ancient Roman lead \cite{romanlead} to further suppress the \gm-rays from the cryostat materials. 
We find the expected environmental muon neutron, and background rates in the ROI to be orders of magnitude smaller than the \alfa~ and \gm~ backgrounds from the experimental apparatus itself \cite{and09,bel10}.
Limited by the background rate, the CUORE projected half-life sensitivity is 1.6 \per 10$^{26}$ y at 1$\sigma$ (9.5 \per 10$^{25}$ y at the 90~\% C.L.) \cite{ale11},  corresponding to an upper limit on the effective Majorana mass in the range 40--100 meV (50--130 meV) depending on the adopted NME calculation \cite{nmes}.  

%\section{CUORE status}
CUORE  is currently in an advanced phase of construction. 
The assembly of the detector, which represented a major effort in the past year and a half has been recently completed and the 19 towers of CUORE are now  stored underground waiting for their installation into the cryostat.
The assembly process represented a true organizational challenge aiming at turning \ca 10000 loose parts into 988 ultra-clean bolometer modules. 
All the operations were conducted in glove boxes under nitrogen atmosphere to minimize oxidization and contamination of radon and radon progenies \cite{cle11}.
The process starts with the gluing of thermistors and heaters to crystals with a robotic system to maintain the uniformity and repeatability of the gluing joints and continue with the mechanical assembly of the instrumented crystals, cleaned copper parts, and PTFE spacers into a tower. 
Two sets of flexible printed circuit board readout cables are then  attached on opposite sides of each tower, and a wire bonder is eventually used to bond the thermistors and heaters to the bonding pads on the readout boards \cite{bro13}.
More details on CUORE detector assembly can be found in \cite{art14}.

\begin{figure}
    \includegraphics[height=0.27\textheight]{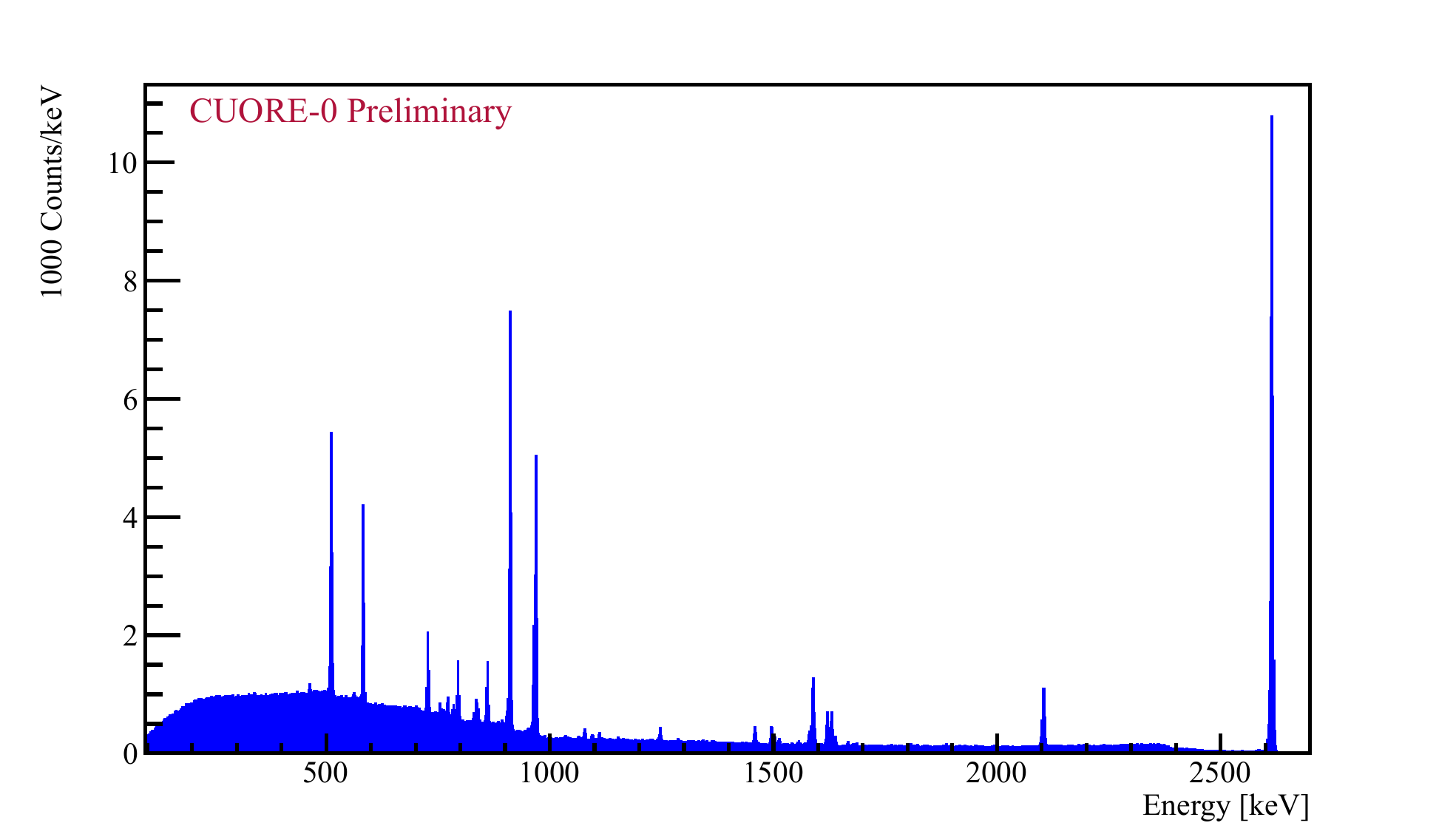}
    \includegraphics[height=0.27\textheight]{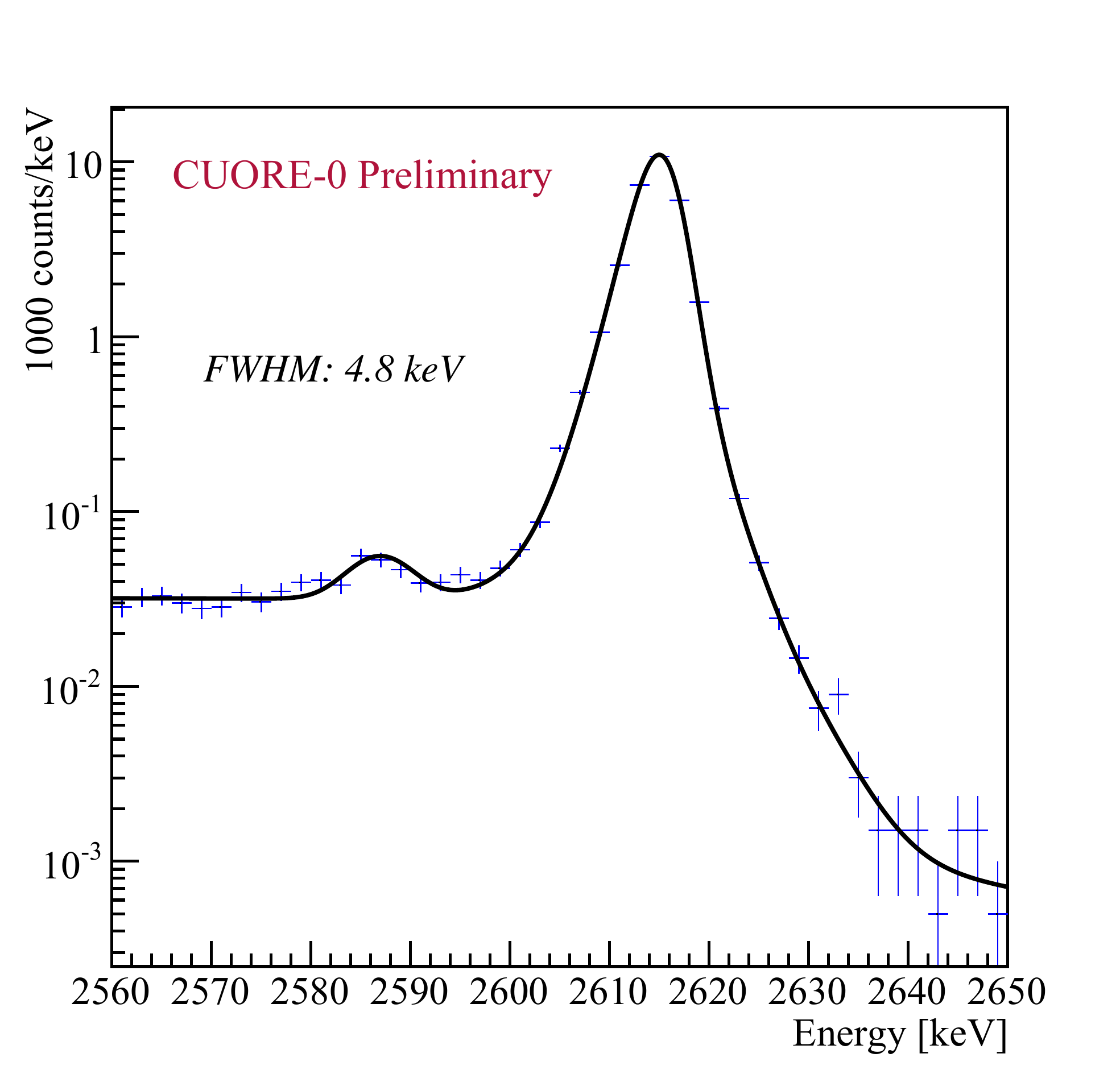}
    \caption{Left: CUORE-0 calibration spectrum (\tdtd external source) from threshold to 2.7 MeV. Events from all the active detectors and all the calibration runs of phase II are included. Right: detail of the \tldo~ line at 2615~keV. The small peak to the left is the Te X-ray escape.}
    \label{fig:calspectra}
\end{figure}

Major efforts are now focused on the commissioning of the cryogenics whose final goal is to cool the bolometer array, almost a ton of material, to 10~mK base temperature. 
This is a very complex system which integrates a large number of sub-systems: a large cryostat (with six nested thermal shields), a cryogen-free cooling system with five pulse-tube coolers and a dilution refrigerator unit, a fast cooling system for pre-cooling, a detector calibration system (DCS), a set of cold (ancient Roman) lead shields, the detector and shield suspensions, the wire readout system and, of course, the CUORE detector \cite{nuc12}. 

The custom-built ${^3}$He/${^4}$He dilution unit (DU) by Leiden Cryogenics has been delivered to LNGS in late Summer 2012 after passing in-house benchmarking. At LNGS it has been tested in a small testing cryostat and successfully reached 5~mK base temperature with a cooling power of 5~W at 12~mK. 

Main goal of the five pulse tubes is to cool down and maintain the system at about 4~K during DU operation. 

On the other hand, the suspensions must accomplish the delicate goal of decoupling the detector from the vibrations of the cryogenic environment while limiting the power injection.

DCS is designed to insert and retrieve 12 radioactive source strings from the 300 K top flange to the bolometer array at 10 mK for energy calibrations every month. The main challenge is to limit the heat load while lowering the source strings through the successive flanges with decreasing temperatures. 
Each Kevlar string carries 25 thoriated tungsten sources crimped on it, evenly separated by 29 mm from each other. 
A motion control system on top of the 300 K flange releases each string to travel down a dedicated guide tube by its own weight. The strings are thermalized at the 4 K flange to avoid over-heating the three inner cryostat vessels. 
%DCS was successfully tested to the 4 K flange together with a commissioning cooldown of the outer cryostat as mentioned earlier.

Given the complexity of the system, a phased commissioning plan has been implemented, characterized by the installation and test of a limited number of items at each stage. 

The commissioning plan started in July 2012 with the outermost three layers of the cryostat (which delimit the two cryostat vacuum chambers) and two of the five pulse tubes needed to cool the system. Vacuum and operational tests of the system took about one year and finished in late spring 2013 with the successful cool-down to 3.5~K \cite{fer14}.

The second phase of the cryogenic system commissioning followed in Summer 2013, with the installation of the remaining three sections of the cryostat, the dilution unit and part of the DCS. 

Vacuum test and the implementation of a good thermometry again took a lot of effort.

The first successful cool-down of the cryostat (run 1) to a base temperature well below the design value of 10~mK (\ca 7~mK) has been recently obtained and we are currently preparing for the next two steps in which the wiring system (run 2), the DCS, the cold shields and the detector suspension plate (run 3) will be installed.
The commissioning plan is expected to finish in May 2015.

Detector integration and the corresponding plan for the installation of the 19 towers into the cryostat are another priority currently addressing a lot of efforts.

In the meanwhile the electronics, the DAQ and the slow control are being finalized and the software development is progressing regularly.
%\subsection{Electronics, data acquisition, and analysis}
The CUORE electronics will read the thermistor voltage signal to computer via the front end electronics, high-precision 18-bit digitizers, and the data acquisition software Apollo. 
Both Apollo and our custom built data analysis software framework Diana have been used extensively for analysis of Cuoricino, R\&D runs \cite{qval}, and more recently CUORE-0. New features and browser-based data quality monitoring are continuously being implemented.

CUORE operation is expected to start at the end of 2015.

%\section{Detector Calibration System}
%\section{cuore:dcs}

\section{3. CUORE-0}
\label{sec:cuore0}

%From CUORE-0 results (sec.~\ref{sec:cuore0}), we evaluate the overall detector energy resolution to be 5.2 keV, based on the FWHM of the 2615 keV peak in the energy spectrum for \BBz searches. This figure reduces to 4.8~keV when considering only the most recent data acquisition phase in which noise contributions from the CUORE-0 cryostat have been improved after a careful refurbishment of the system during fall 2013. It should be also noted that CUORE-0 has been running at a base temperature range 13–15 mK, higher than expected CUORE base temperature at 10 mK. In our R\&D runs, when the bolometers were operated below 13 mK, we consistently achieved energy resolution below 5 keV \cite{art14b}.

CUORE-0 is a CUORE-style tower built using the same materials, assembly devices and procedures developed for CUORE.
% (Fig.~\ref{fig:cuoredet}).
It is therefore identical in all the aspects to the 19 towers of CUORE (52 \teod crystals, 39~kg of \teod mass and 11 kg of \tect). 
Actually, the main goal of CUORE-0 is to provide a proof of concept of the CUORE detector in all stages, including the operation of the DAQ and analysis frameworks.
Given the sizable mass and the anticipated background level, CUORE-0 is operating as an independent \BBz experiment while CUORE is under construction.

\begin{figure}
\includegraphics[height=0.32\textheight]{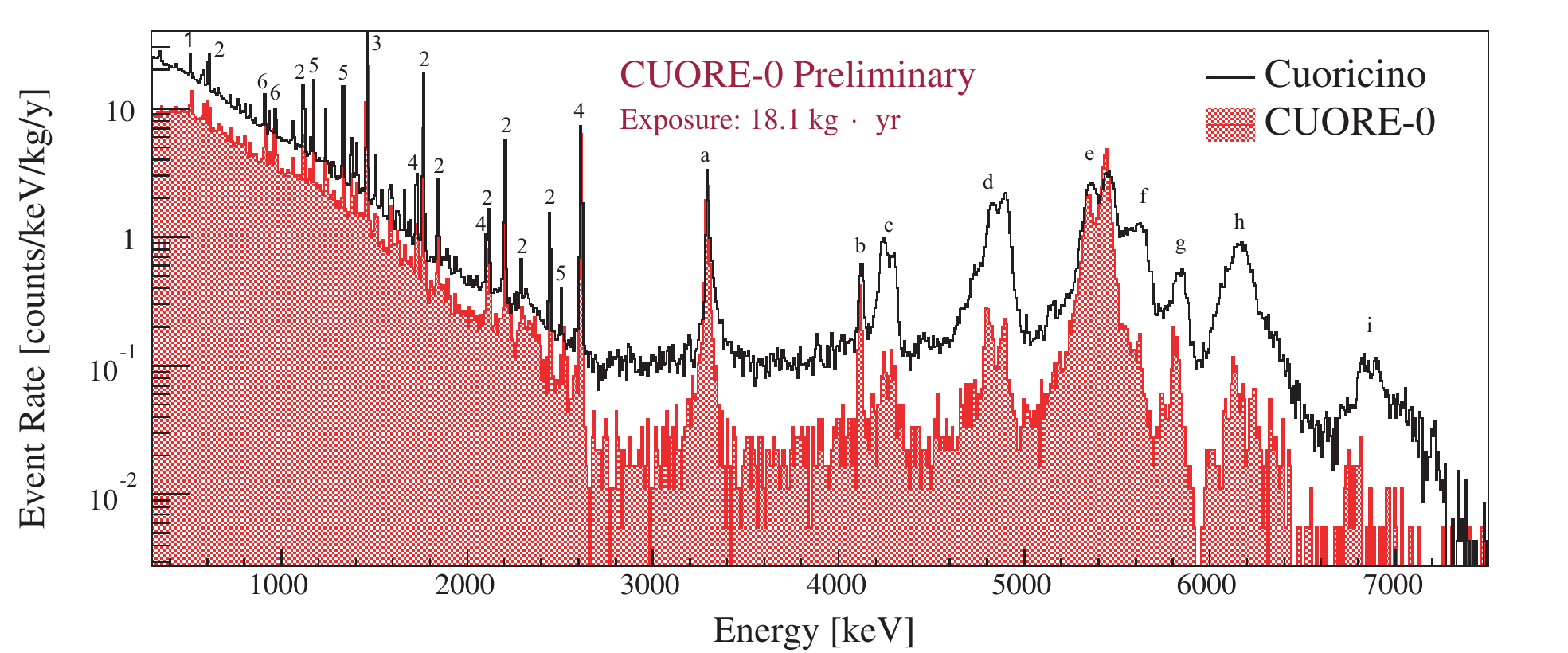}
\caption{Cuoricino (line) and CUORE-0 (shaded) background spectra comparison. 
Only events with a single crystal hit are considered (anti-coincidence mode). Background rates are clearly reduced in CUORE-0 by factors \ca 2 and \ca 6 in the \alfa~(E>2.7~MeV) and \gm~dominated regions, respectively. In the \gm~region, labeled lines are due (1) e+e- annihilation, (2) \bdq, (3) \kq, (4) \tldo, (5) \cs~and (6) \advo~mainly from the cryostat materials.  In the \alfa region they come from (a) \pcn, (b) \tdtd, (c) \udto, (d) \tdt~and \rdvs, (e) \pdd, (f) \tdvo~and \rdvd, (g) \rdvq, (h) \pddo~and (i) \pds~from \teod crystals or surface radioactive contaminations of the detector structure materials.
}
\label{fig:bkgspectra}
\end{figure}

CUORE-0 is operated in the same cryostat that previously hosted Cuoricino, in the Hall A of LNGS \cite{and11}]. 
CUORE-0 maintains an operating temperature of about 13-15 mK. At this temperature the typical signal amplitude is 10--20 $\mu$K/MeV, with typical rise and decay times of 50 and 250 ms, respectively.
The analog read-out of the thermistor is performed using the same electronics that was used for Cuoricino \cite{arn03}. 
The signals are first amplified, filtered by 6-pole active Bessel filter \cite{arn10} and then fed into an 18-bit National Instrument PXI analog-to-digital converter (ADC). 
The filter cutoff and the ADC sampling frequency are set to 12 Hz and 125 Hz, respectively. 
The data are then processed with Apollo, the data acquisition software developed for CUORE. 
A CUORE-0 {\it dataset} consists of 3-4 weeks of low background data taking preceded and followed by 2--3 days of calibration runs. To calibrate the detector, we insert two thoriated tungsten wires between the outer vacuum chamber of the cryostat and the external lead shield. During the offline analysis, we calibrate each channel separately over the energy range 511 to 2615 keV using the  lines from the daughter nuclei of \tdtd.

The raw data from the bolometers are processed offline with Diana, the analysis software suite developed for CUORE. 
The data reconstruction steps are the same as in the Cuoricino analysis \cite{and11}, but we have improved the automation and robustness of many of the algorithms in anticipation of scaling to CUORE.

The offline data production begins by evaluating the signal amplitude of each waveform using the matched filter described in \cite{rad67,gat86}. 
Noisy pulses are then identified and discarded as a result of the comparison with a set of template signals.
Gain drifts caused by thermal variations are then corrected by exploiting the Joule heater attached to each crystal. 
Calibration spectra are analyzed to extract the correct calibration function of each detector. 
The final step of the data production, is to evaluate the time coincidences between events on different crystals which are later used in the anti-coincident analysis.
Since the majority of \BBz events are expected to be fully contained in a single crystal, only single crystal events are selected.

\begin{figure}
    \includegraphics[height=0.25\textheight]{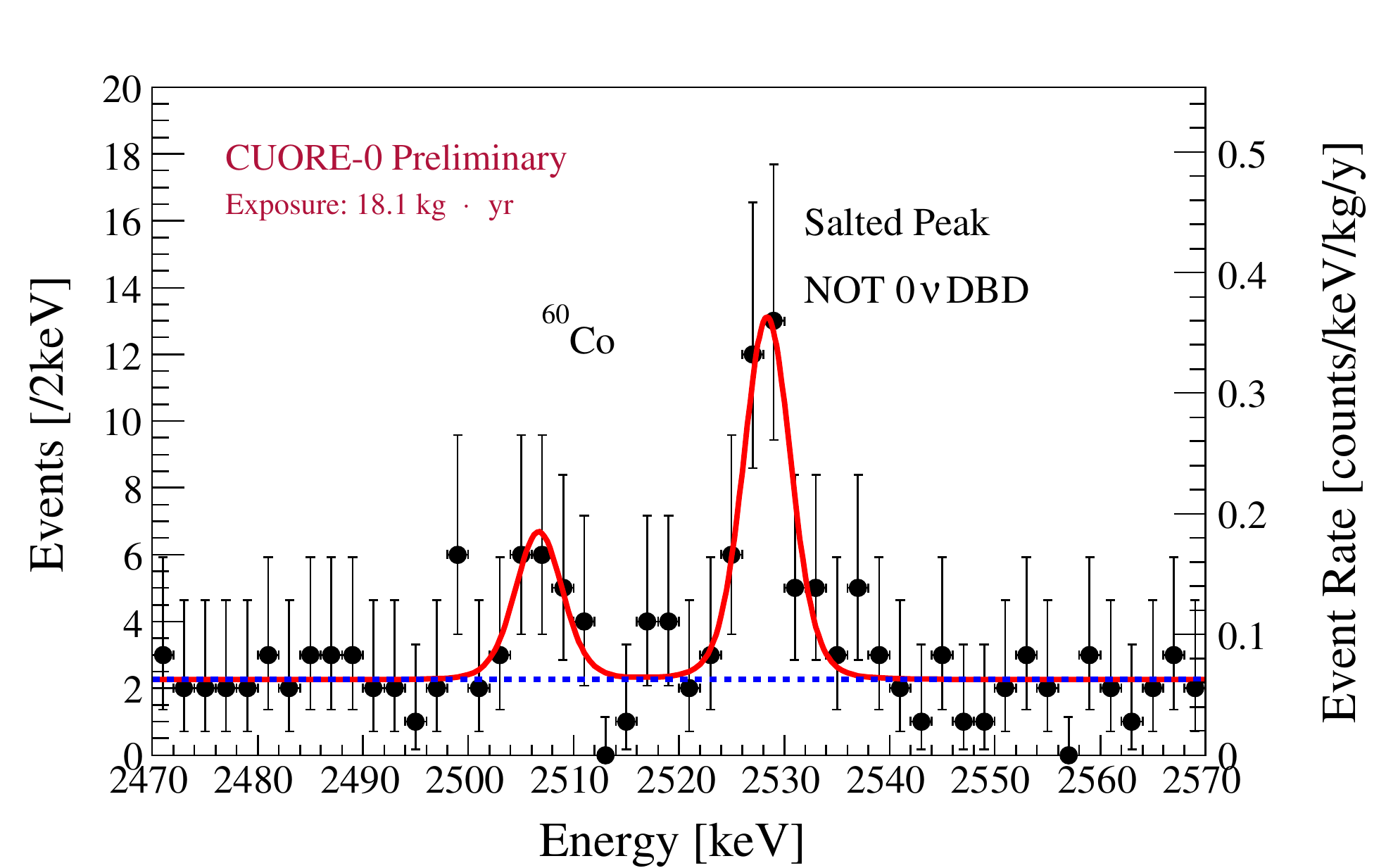}
    \includegraphics[height=0.25\textheight]{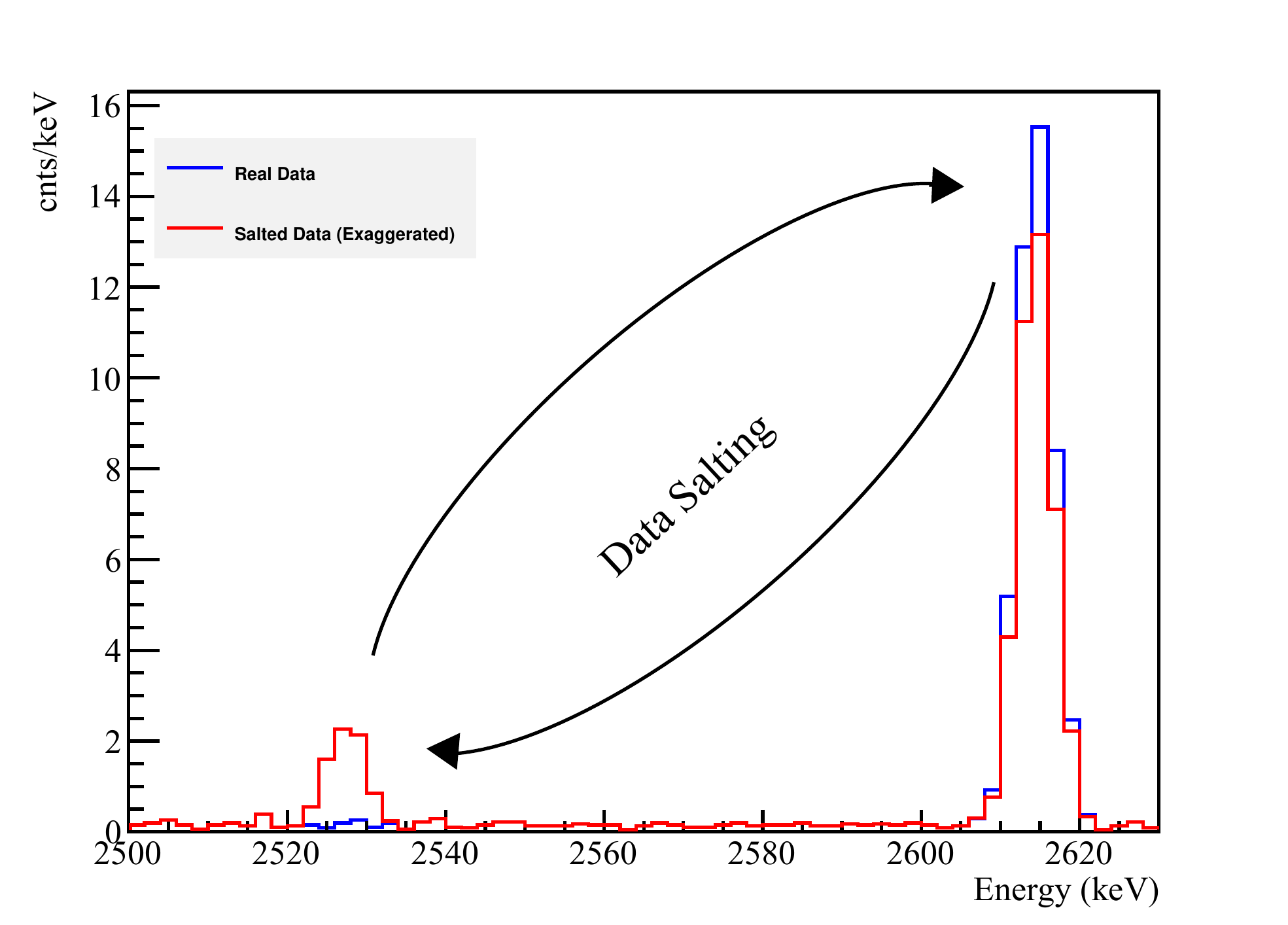}
    \caption{Left: CUORE-0 \BBz energy region of interest. The peak at 2506 keV is due to the sum of the two \gm's from \cs. The peak at 2528 keV is
	the salted \BBz peak (see text). Right: CUORE-0 blinding concept based on the exchange of an unknown small fraction of events between ROI and \tldo~ line.}
    \label{fig:bb0}
\end{figure}

The overall detection efficiency in this analysis is 77.6 \pom 1.3~\%, which includes the shape, anti-coincidence and containment efficiencies.
Efforts towards optimized cut efficiencies are ongoing.
The \BBz ROI is kept blinded using a procedure of ``data salting'' where an unknown (\ca  1--3~\%) fraction of events from the 2615 keV line are randomly selected and moved to the \BBz ROI and vise-versa.  Since there are more events in the 2615 keV peak than the ROI, this process produces a fake peak in the ROI  (Fig.~\ref{fig:bb0}). 
The advantage of this blinding procedure is that it preserves the energy spectrum in the ROI, and allows one to test the fitting algorithms that will be used after
unblinding.

More details on the CUORE-0 data taking and analysis procedures can be found in \cite{art14c,vig13}.

%--- Results
CUORE-0 data taking started in March 2013, though in non optimal conditions, after a troubled preparation of the detector which identified unexpected flaws in the CUORE preparation procedures that were then fixed. 
Main goal of the following phase of data taking was to measure the performance parameters while collecting statistics for evaluating the background level near the \BBz ROI. The first CUORE-0 data were published in the Summer 2013 \cite{art14c,vig13}. A long system maintenance then followed during which a full section of the dilution unit condensing line was completely replaced and other sections were refurbished.
The result was a new phase (named ``II'') of data taking, started in November 2013 and characterized by a terrific improvement of the system performance both in terms of noise and stability.
Figure \ref{fig:calspectra} (left) shows the sum of all the calibration spectra collected so far, during phase II. All active channels are included. 
The measured FWHM energy resolution at the 2615~keV line of \tldo~ is 4.8~keV, a sizable improvement with respect to phase I result \cite{vig13} and better than the anticipated design goal of CUORE. 
The total background exposure includes both phase I and II and amounts to 18.1 kg \per y (6.2 kg \per y of \tect ). 
In the low background spectrum (Fig.~\ref{fig:bkgspectra}), the peaks from \tldo, \kq~and \cs~are attributed to contamination in the cryostat and the structure materials, while the peaks from \bdq~are attributed to \rdvd~in the air around the cryostat during the initial runs. On the other hand, the peaks in the \alfa~region are attributed to bulk contaminations of the \teod crystals, while the other structures are attributed to surface contamination of the \teod crystals and/or the close materials.
The energy resolution evaluated on the \tldo~ line at 2615~keV is 5.2~keV FWHM, half-way between the improved performance of phase II and the 5.7~keV measured during phase I \cite{art14c}.
The continuum from 2.7 to 3.9 MeV, excluding the only clearly visible peak ascribed to \pcn, is attributed to degraded particles that deposit only a fraction of their energy in the crystal and the rest in inactive materials. 

This \alfa~continuum extends down into the ROI and was a significant background for Cuoricino, 0.110 \pom 0.001 \ckky. 
The value currently measured in CUORE-0 is 0.020 \pom 0.001 \ckky. 
The factor of \ca 6 reduction from Cuoricino proves the success of the radio-purity protocols that were implemented for the CUORE production and assembly.
The energy spectrum in the \BBz ROI is shown in Fig.~\ref{fig:bb0}. The false peak produced by the data salting is clearly evident. The fit consists of a flat background with two gaussians, one for the 2506 keV \cs~sum peak and one for the 2528 keV \BBz peak. 
The flat background rate in the ROI is measured to be 0.063 \pom 0.006 \ckky. 
The excess over the \alfa~background stated above is attributed to scattered 2615 keV \gm's originating from the cryostat. 
This excess is consistent with the rate similarly observed in Cuoricino, which was run in the same cryostat.

CUORE-0 data taking is expected to continue until the start of CUORE. 
Data unblinding is however expected when the CUORE-0 sensitivity will surpass the Cuoricino limit. This is expected in spring 2015.

\section{4. Prospects}
\label{sec:ihe}

Successful cryogenic operations of CUORE, as well as the CUORE experience in ultra-clean assembly of bolometers and a cryostat system, are critical for demonstrating the viability of future bolometric experiments.
However, CUORE sensitivity is mainly limited by background.
Based on the background model built over a number of dedicated R\&D measurements, but especially based on Cuoricino and CUORE-0 data, the main issue are surface radioactive contaminations, characterized by the emission of \alfa~and $\beta$/\gm's in the MeV region.
Particle identification is therefore the main emphasis on bolometers future application in rare event searches. 
%While our main focus is CUORE construction and operation, active R\&D programs are already indicating that the background suppression by two orders of magnitude in bolometers is viable.
To this end, additional detection channels are needed, since the absorber does not respond differently for an energy release of different particle types. 
To distinguish between \BBz electrons and \alfa~particle background, light emission, either from Cherenkov radiation or scintillation light can be exploited, where the auxiliary light detector is usually another bolometer facing the main bolometer. 
Alternative methods based on the identification of surface interactions have also been devised.
%The light detector nominally consists of a thin germanium/silicon wafer as absorber and a thermal sensor of the same type as the main bolometer.
This new generation of bolometric experiments aim at reaching a sensitivity to explore the inverted neutrino mass hierarchy region, at an effective Majorana mass \mbb~\ca 10 meV \cite{ihe}. 

With the current support from the American and European funding agencies, a number of active R\&D programs are already investigating the cost and purity of bolometric crystals highly enriched in \tect~ as well as other potential isotopes (\sod, \mc, and \ccs), studying background rejection with scintillation, Cherenkov radiation, ionization, and pulse-shape discrimination, testing novel materials and sensor technologies.

Indeed, Cherenkov light radiation generated by electrons with energy larger than 50 keV \cite{tab10} can provide a definite discrimination of \alfa~backgrounds and can be exploited in all practical cases.
Actually Cherenkov light emission in \teod (which does not scintillate) has been detected using a secondary light detector  \cite{bee12} but more efficient light collection schemes and light detectors with lower threshold and better energy resolution are needed. 
Future directions include new thermal sensors on the light detector, such as TES \cite{ang12} and MKIDs \cite{did14}, and Luke effect \cite{isa12} enhanced bolometers.

When releasing the request to use \teod absorbers, scintillating bolometers are a natural choice. They operate similarly to the Cherenkov case, but with a larger light yield. 
The particle identification comes then from the different light yield for different particle type. 
For scintillating crystals such as \zs, \ct, and \zm, the light yield difference between possible \BBz electrons and \alfa~particles have been measured \cite{pir06} and the background suppression concept demonstrated. It is worth noticing that the natural isotopic abundance of \sod, \ccs~and \mc~are all less than 10~\% so that isotopic enrichment is mandatory.

Experience with the CUORE assembly will allow us to further refine and optimize the process of putting together and operating a 1-ton scale detector. 
Once CUORE is operational, we expect the vigor of these R\&D activities to ramp up, with the goal of preparing a full proposal for an upgraded bolometric experiment
with O(10~meV) Majorana mass sensitivity on the timescale of 2016.
In fact, with a 90~\% C.L. half life sensitivities of the order of 10$^{27}$ y, a CUORE-like experiment with the substantial background reduction provided by additional detection channels would be able to reach a sensitivity on \mbb~of the order of 5-15~meV, and explore the inverted hierarchy region of neutrino masses \cite{ihe}.

\section{5. Conclusions}

We described the physics reach and current construction status of the CUORE experiment. 
Start of data taking is expected in 2015. 
With excellent energy resolution and large isotope mass, CUORE is one of the most competitive \BBz experiments under construction.

With an isotope mass of 11 kg, a background index of 0.063 \pom 0.006 \ckky~and an energy resolution of 5.2 keV FWHM, CUORE-0 is the most sensitive experiment searching for the \BBz of \tect. 
It is expected to surpass the Cuoricino sensitivity in about 1 year of live time.
The background in CUORE-0 is dominated by \gm's from the old Cuoricino cryostat. 
The measured \alfa~background index of 0.020 \pom 0.001 \ckky~validates the background reduction techniques developed for CUORE. 
Projecting the observed background to CUORE, where the cryostat contaminations are expected to be negligible, the target background index of 0.01 \ckky~in the ROI seems achieved.
With this background CUORE is expected to achieve a sensitivity to the \BBz half-life of \tect~ of \ca 10$^{26}$ y \cite{ale11}.

Future bolometer program can explore the inverted mass hierarchy region to \mbb \ca 10~meV by utilizing additional detection channels to further reduce background, which is presently the major limitation to the CUORE sensitivity.
Moved by the progress in the construction of CUORE and the already available R\&D results, a group of interest is presently forming to propose this technology as one of the most promising future programs to search for \BBzn.

\section{Acknowledgements}
The CUORE Collaboration thanks the directors and staff of the Laboratori Nazionali del Gran Sasso and the technical staff of our laboratories. This work was supported by the
Istituto Nazionale di Fisica Nucleare (INFN); the National Science Foundation under Grant Nos. NSF-PHY-0605119, NSF-PHY-0500337, NSF-PHY-0855314, NSF-PHY-0902171, and NSF--PHY-0969852; the Alfred P. Sloan Foundation; the University of Wisconsin Foundation; and Yale University.
This material is also based upon work supported by the US Department of Energy (DOE) Office of Science under Contract Nos. DE-AC02-05CH11231 and DE-AC52-07NA27344; and by the DOE Office of Science, Office of Nuclear Physics under Contract Nos. DE-FG02-08ER41551 and DEFG03-00ER41138. This research used resources of the National Energy Research Scientific Computing Center (NERSC).

\bibliographystyle{aipprocl}

\end{document}